\documentclass[aip, apl , preprint]{revtex4-1}

\usepackage{docs}%
\usepackage{bm}%
\usepackage{subcaption}
\usepackage[pdftex]{graphicx}
\graphicspath{{figures/}}	
\usepackage{float}
\usepackage[colorlinks=true,linkcolor=blue]{hyperref}%
\expandafter\ifx\csname package@font\endcsname\relax\else
\expandafter\expandafter
\expandafter\usepackage
\expandafter\expandafter
\expandafter{\csname package@font\endcsname}%
\fi
\usepackage{amsmath}
\usepackage{gensymb}	

\begin{document}

	\title{Low-loss integrated photonics for the blue and ultraviolet regime}
	\author{Gavin N. West}
	\affiliation{Department of Electrical Engineering and Computer Science, Massachusetts Institute of Technology, Cambridge, Massachusetts, 02139, USA}
	\author{William Loh}
	\affiliation{Lincoln Laboratory, Massachusetts Institute of Technology, Lexington, Massachusetts 02421, USA}
	\author{Dave Kharas}
	\affiliation{Lincoln Laboratory, Massachusetts Institute of Technology, Lexington, Massachusetts 02421, USA}
	\author{Cheryl Sorace-Agaskar}
	\affiliation{Lincoln Laboratory, Massachusetts Institute of Technology, Lexington, Massachusetts 02421, USA}
	\author{Karan K. Mehta}
	\affiliation{Department of Electrical Engineering and Computer Science, Massachusetts Institute of Technology, Cambridge, Massachusetts, 02139, USA}
	\author{Jeremy Sage}
	\affiliation{Lincoln Laboratory, Massachusetts Institute of Technology, Lexington, Massachusetts 02421, USA}
	\author{John Chiaverini}
	\affiliation{Lincoln Laboratory, Massachusetts Institute of Technology, Lexington, Massachusetts 02421, USA}
	\author{Rajeev J. Ram}
	\affiliation{Department of Electrical Engineering and Computer Science, Massachusetts Institute of Technology, Cambridge, Massachusetts, 02139, USA}
	
	\maketitle

	We present a low-loss integrated photonics platform in the visible and near ultraviolet regime. 
	Fully-etched waveguides based on atomic layer deposition (ALD) of aluminum oxide operate in a single transverse mode with $<$3 dB/cm propagation loss at a wavelength of 371 nm.
	Ring resonators with intrinsic quality factors exceeding 470,000 are demonstrated at 405 nm, and the thermo optic coefficient of ALD aluminum oxide is estimated to be $2.75\times10^{-5}$ [RIU/$^\circ$C]. 
	Absorption loss is sufficiently low to allow on-resonance operation with intra-cavity powers up to at least 12.5 mW, limited by available laser power. 
	Experimental and simulated data indicates the propagation loss is dominated by sidewall roughness, suggesting lower loss in the blue and UV is achievable.

	\section{Introduction}
	The success of silicon photonics in telecommunications has lead to  the application of nano-scale photonics in a variety of fields including computing, nonlinear optics, quantum information processing, and biochemical sensing \cite{POEM,nonlinear_silicon,KaranNatureNano,linear_quantum_optics,SiRingBiosensor}. 
	Compact device footprints and an ability to leverage the same manufacturing techniques employed in the semiconductor industry are strong incentives both for systems designers and in applications where low cost is necessary. 
	Label-free biosensors, optical interconnects for computers and datacenters, integrated lasers with III-V gain media, and phased arrays consisting of thousands of elements have all been demonstrated using the same basic silicon photonic technology \cite{SiRingBiosensor,SiRingBiosensor2,POEM,BowersLasers,WattsPhasedArray}.
	However, with a bandgap at 1.1 um, silicon is unsuitable for applications which require visible or ultraviolet light, such as optogenetics \cite{optogenetics_pecvd_sin, optogenetics_lpcvd_sin}, protein sensing \cite{ForesterResonanceUV,TryptophanUV}, and atom-based sensing, time-keeping, and information processing \cite{ atom_cooling,optical_clocks,yb_lattice_clock,neutral_atom_gate}. 
	A straightforward way of bypassing this limitation is to use silicon nitride, commonly integrated alongside silicon, which has transparency into the visible. 
	Waveguide platforms based on silicon nitride are quite mature, particularly for red and near infrared (NIR) wavelengths. 
	Less progress has been made for devices operating in the blue and near ultraviolet (UV, NUV). 
	This is predominantly due to the high material absorption ($>$20 dB/cm) that begins in the low 400 nm wavelength range \cite{LL_photwest}.

	The most common alternative to silicon nitride for ultraviolet photonics are the III-V nitride materials, particularly aluminum nitride (AlN) and aluminum-gallium nitride alloys (AlGaN) \cite{DirkSorefNitrides}. 
	AlN has a bandgap corresponding to $\lambda \sim$ 200 nm and exhibits second-order nonlinearities, making it attractive for integrated nonlinear optics and electrical tuning of resonant structures. 
	Early demonstrations of ultraviolet waveguides in AlN suffered extremely high loss (389 dB/cm at a wavelength $\lambda =$ 450 nm) due to a combination of bulk (polycrystalline) material loss and high sidewall roughness \cite{first_uv_aln}.  
	More recent work using nanocrystalline AlN has brought the loss coefficient down by an order of magnitude ($\sim$50 dB/cm at $\lambda =$ 405 nm) but in a regime where device size is restricted to sub-centimeter or sub-millimeter lengths when significant power attentuation is a concern \cite{JeffLuAlN}. 
	At the time of writing, initial results with single crystal AlN grown on sapphire by metal-organic CVD have been reported with losses as low as $\sim$8 dB/cm at $\lambda = $390 nm \cite{hongtang_aln}. 
	This single-crystal growth is restricted to sapphire substrates, where the lattice mismatch is small, but reduces the strong scattering coefficients of polycrystalline AlN at short wavelengths, and exhibits lower impurity concentration. 
	Single crystal bulk AlN substrates have recently become available \cite{hexatech_aln,aln_sc_material_loss_2}, but still suffer from high defect density and are limited to 25mm wafers. 
	
\begin{figure}[H]
			\centering {
				\phantomsubcaption\label{fig:fig1:a}
				\phantomsubcaption\label{fig:fig1:b} \phantomsubcaption\label{fig:fig1:c}  
					}
			\includegraphics{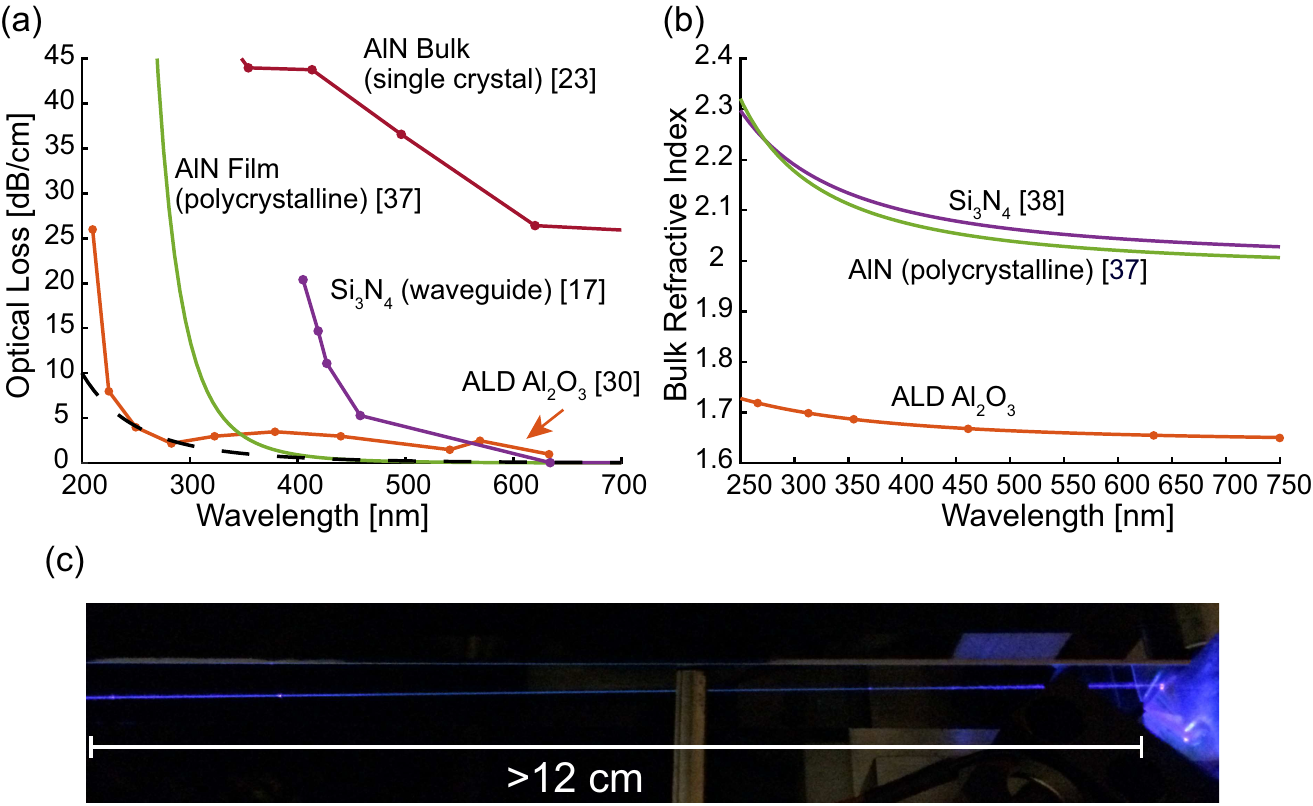}
			\caption{
			\subref*{fig:fig1:a} Literature values of loss in various materials used for UV-vis integrated photonics \cite{Aslan_UV,aln_poly_material_loss,aln_sc_material_loss_2,LL_photwest}.   
			The dashed line represents $\propto 1/\lambda^4$ scattering scaled to 0.1 dB/cm at $\lambda =$ 633 nm. 
			\subref*{fig:fig1:b} Refractive index dispersion for each material, from literature (Si$_3$N$_4$ and AlN) \cite{luke_sin_index,aln_poly_material_loss}, and ellipsometry with Cauchy fit (Alumina).
			\subref*{fig:fig1:c} Prism-coupled $\lambda =$405 nm light propagating $>$ 12 cm in a 200 nm thick ALD Al$_2$O$_3$ film. }
			\label{fig:fig1}
		\end{figure}

	As an alternative, amorphous aluminum oxide (alumina) has an electrical bandgap between 5.1 and 7.6 eV ($\lambda \sim$163-243 nm) depending on the deposition mechanism \cite{bandgap_51,bandgap_62,bandgap_70,bandgap_76} and has been used in both slab waveguides at telecommunications wavelengths \cite{alox_slab_telecom} and as a back-end integrated waveguide in CMOS \cite{amir_ieee_sum}. 
	Optical characterization has shown low transmission loss down to a wavelength of 220 nm when deposited by atomic layer deposition (ALD) \cite{Aslan_UV}. 
	Fig. \ref{fig:fig1}\subref{fig:fig1:a} compares reported loss in AlN, Si$_3$N$_4$, and alumina, demonstrating the large transparency of alumina.
	Our measured refractive index, $n$, of alumina is 1.65-1.72 in the visible-NUV spectrum, suitably higher than $n =$1.45-1.49 of silicon dioxide typically used for cladding. 
	Due to its lower refractive index, as compared with silicon nitride or aluminum nitride ($n\sim 2-2.3$ , Fig. \ref{fig:fig1}\subref{fig:fig1:b}), alumina will experience less strong sidewall scattering for equivalent roughness and the minimum device feature size (scaling roughly as $\lambda/n$) will be larger \cite{lipson_book,haus_barwicz}. 
	Many structures are already at the edge of the $\sim$ 100 nm resolution of deep-UV photolithography, making the larger device size a boon. 
	The tradeoff is that lower confinement results in higher bending loss for a given bend radius.

	The ALD alumina process based on trimethylaluminum (TMA) and water, being developed for high-k gate dielectrics by the semiconductor industry, is one of the best-understood ALD processes. 
	Films are extremely uniform, conformal, and exhibit low defect densities \cite{ald_alox_review}. 
	Additionally, the layer-by-layer deposition nature of ALD lends itself to strict tolerances, assisting repeatability in run-to-run device behavior. 
	The common roadblock to fabricating nanophotonic devices with alumina is patterning--reactive ion etch (RIE) chemistries typically etch at slow rates with high sidewall roughness \cite{AlOx_Etched_Rib,AlOx_etched_er_amplifier}, and available wet etches do not provide the anisotropy required for integrated photonic structures.
	As a result, to our knowledge there has been no demonstration of a fully-confined alumina waveguide at short wavelengths.
	Previous examples of alumina waveguide structures have employed simple films, partial etches, or pre-etched templates into which material is deposited \cite{alox_slab_telecom,AlOx_Etched_Rib,AlOx_etched_er_amplifier,watts_alox_er_laser,amir_ieee_sum}. 
	These methods do not provide the same geometric control or modal confinement as fully etched alumina. 
	Here we realize directly-etched waveguides with steep, smooth sidewalls which exhibit low loss at short wavelengths.

	\pagebreak
	\section{Material Characterization}
	
	Alumina films were grown on bare silicon and on thick (3.2 $\mu m$) SiO$_2$ on silicon using TMA and water precursors (Oxford Opal reactor). 
	This reaction is known to leave trace carbon impurities, which can be reduced by increasing the reaction temperature or by annealing \cite{alox_carbon_content}. 
	Further, the refractive index tends to increase at higher deposition temperatures. 
	We chose a 300\degree C growth temperature, limited by reactor temperature limits, to promote defect-free deposition, and measured a growth rate of $\sim$1 angstrom per cycle. 
	The refractive index increases from 1.65 at 633 nm to 1.72 at 260 nm, as measured by spectroscopic ellipsometry (KLA-Tencore UV1280), shown in Fig. \ref{fig:fig1}\subref{fig:fig1:b}. 
	Film surface roughness measured by an atomic force microscope (AFM) was 0.34 nm RMS over a 5$\mu$m$\times$5$\mu$m area.
	Optical loss in the films was measured using the prism coupling method (Metricon) at both 633 nm and 405 nm, at less than $<$0.3 dB/cm, with measurement sensitivity limited by stray light.
	 \ref{fig:fig1}\subref{fig:fig1:c} shows an un-etched alumina film in the Metricon, propagating 405 nm light.
	Further increases in the refractive index can be realized by high temperature annealing, however this causes formation of dense polycrystalline $\gamma$-phase Al$_2$O$_3$ above 800\degree C \cite{bandgap_70}. 
	Alumina annealed at 900\degree C and 1100\degree C is polycrystalline, as seen in Fig. \ref{fig:fig2}\subref{fig:fig2:d}, and exhibits optical losses $>$20 dB/cm.

	\section{Waveguide Fabrication}

	Choosing an appropriate mask material is paramount as alumina is attacked by most strong acids and bases \cite{AlOx_AcidBaseEtch}. 
	This makes post-etch mask removal difficult, so the mask material must be compatible with the application. 
	Here we use plasma-enhanced chemical vapor deposition (PECVD) SiO$_2$ as a hard mask, for two reasons: it is relatively resistant to the chlorine-based RIE chemistries we use to etch alumina, and the remaining mask is transparent in the blue and NUV spectrum and thus can be left in place.
	
	For all samples, 3.5$\mu m$ PECVD oxide was grown (Novellus Sequel) on bare 200 mm silicon wafers followed by a 4 hour, 1100\degree C anneal and chemical-mechanical polishing (CMP) to a final thickness of 3.2 $\mu$m. 
	The alumina waveguide core layer was deposited at 300\degree C in an Oxford Opal ALD reactor with TMA and H$_2$O precursors to a target thickness of 100 nm. 
	For the hard mask, a 200 nm thick SiO$_2$ layer was then deposited via PECVD at 400\degree C. 
	Patterning of the hard mask was done on an ASML 5500 193nm stepper using 300 nm-thick JSR resist with a thin anti-reflection coating (ARC). 

\begin{figure}[H]
			\centering{
				\phantomsubcaption\label{fig:fig2:a}
				\phantomsubcaption\label{fig:fig2:b} 
				\phantomsubcaption\label{fig:fig2:c} 
				\phantomsubcaption\label{fig:fig2:d}
						}
			\includegraphics{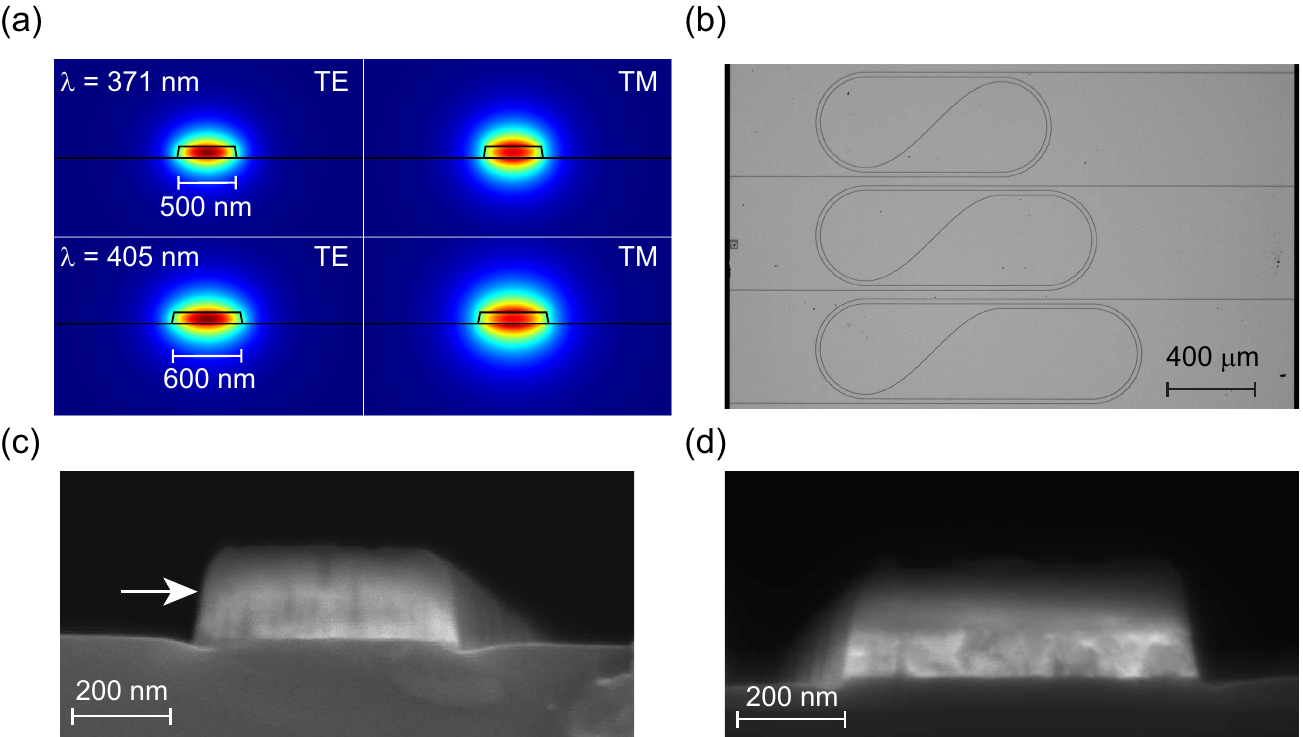}
			\caption{
				\subref*{fig:fig2:a} Simulated mode profiles of the TE and TM modes with 500 and 600 nm wide waveguides for $\lambda =$ 371 and 405 nm, respectively, for 100 nm core height. 
				The mode effective indices are approximately 1.50. \subref*{fig:fig2:b} Example paperclip structures used to measure propagation loss. 
				The dark edge on each end is the etched facet defined for butt coupling. 
				\subref*{fig:fig2:c} An example etched alumina waveguide. 
				The white arrow denotes the line between alumina and remaining SiO$_2$ hard mask.  
				Top-down AFM measurements of the waveguide sidewall give an RMS roughness of 1.4 nm and correlation length of 29 nm. 
				\subref*{fig:fig2:d} The polycrystalline formations after a 900\degree C anneal are clearly visible under SEM.
				}
			\label{fig:fig2}
		\end{figure}

	After development the pattern was transferred to the SiO$_2$ hard mask with C$_4$F$_6$/CF$_4$ RIE. 
	Any remaining ARC/photoresist was removed under low-bias oxygen RIE. 
	Failure to remove the residual resist was observed to result in poor alumina etching quality. 
	The alumina layer was etched in an inductively coupled plasma (ICP) RIE (Applied Materials Centura) using a BCl$_3$/Ar chemistry in 1:2 ratio. 
	Increasing the ICP/bias power ratio (near tool limits), creating a more energetic plasma and reducing the effects of sputter etching, was found to improve etch rate and selectivity. 
	The resulting sidewall angle is $\sim$80 degrees with respect to horizontal.
	An example etched ridge is shown in Fig. \ref{fig:fig2}\subref{fig:fig2:c}, the interface between the alumina core and SiO$_2$ cladding is denoted by the white arrow. 
	3 $\mu$m PECVD oxide is deposited as a cladding, with a 400\degree C anneal in nitrogen.

\pagebreak 
	Waveguide structures were patterned with three widths, from 400 to 600 nm, chosen to be below the single-mode cutoff at $\lambda = 405$ nm. 
	Simulated mode profiles are shown in Fig. \ref{fig:fig2}\subref{fig:fig2:a}.
	At $\lambda = 371$ nm the 600 nm width is expected to be multimode.
	A final etch was used to define a facet for butt coupling, with no additional polishing.

	\section{Testing and Analysis}
	
	\subsection{Waveguide Loss Measurements}

	Propagation loss was measured using paperclip structures of increasing length (Fig. \ref{fig:fig2}\subref{fig:fig2:b}), with a maximum 7 cm path length differential. 
	The entire 200 mm wafer was patterned with dozens of copies of the paperclips and rings.
	No significant difference in loss was observed from die to die.
	Bends in the paperclips are a minimum of 400 $\mu$m diameter, such that radiation loss is negligible, and each paperclip has an identical number of bends.
	Loss was extracted as the slope of a linear fit of the logarithm of the output power versus waveguide length. 
	
\begin{figure}[H]
		\centering{
			\phantomsubcaption\label{fig:fig3:a}
			\phantomsubcaption\label{fig:fig3:b}
				}
		\includegraphics{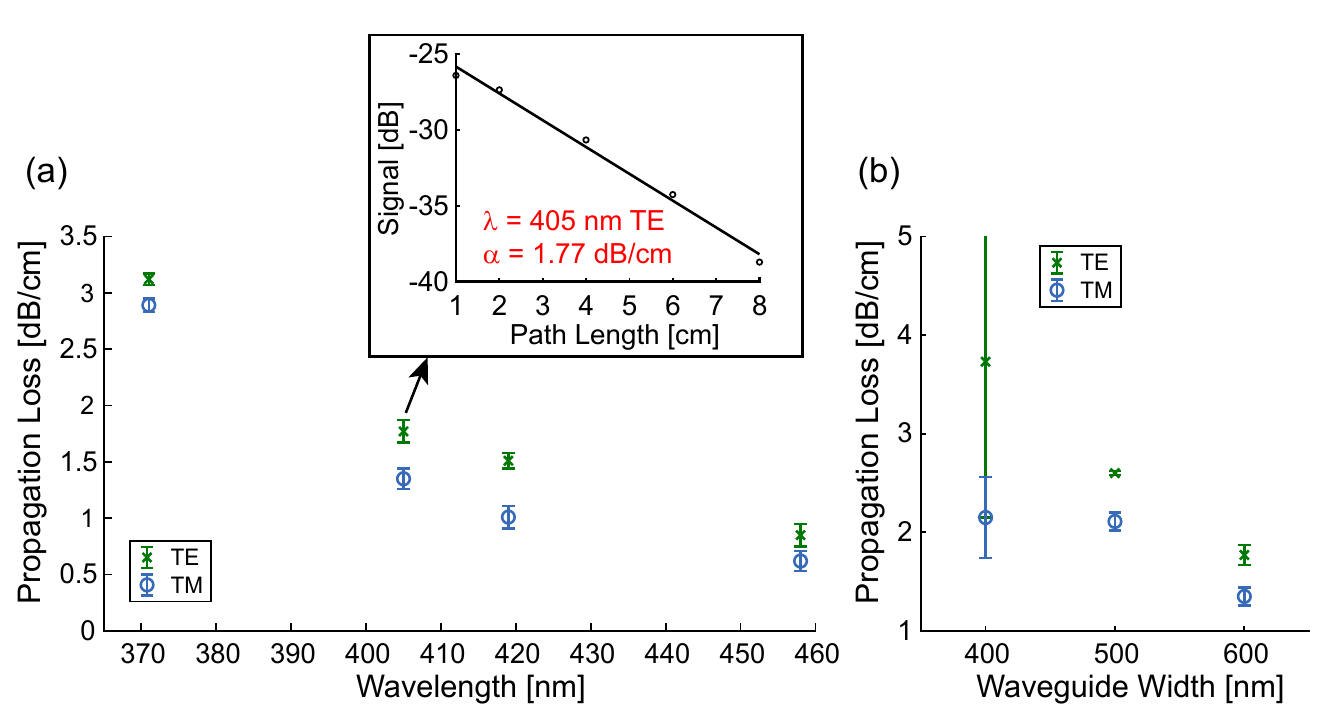}
		\caption{
			\subref*{fig:fig3:a} Propagation loss in the blue and NUV spectrum.
			Inset: An example curve fit from measured data, for TE polarized 405 nm light in a 600 nm wide waveguide. 
			\subref*{fig:fig3:b} Dependence of propagation loss on waveguide width at $\lambda =$ 405 nm. 
				}
		\label{fig:fig3}
	\end{figure}

	Laser sources consisted of single-frequency diode lasers emitting at 371 nm, 405 nm, 419 nm, and 458 nm, coupled into polarization maintaining single mode fiber.
	Input and output coupling was achieved either with a 40X microscope objective or by direct butt coupling of cleaved fiber to the etched facet. 
	Polarization was controlled using a half wave plate positioned before fiber launch.
	Piezo-driven 3-axis stages were used to align input/output fiber or objectives with the waveguide core.
	Variations in this coupling can occur from roughness introduced by the facet etch and misalignments of the input mode, and appear in the uncertainty of the fitted slope.
	This uncertainty was generally quite small, $<0.1$ dB/cm slope error and $<0.4$ dB error in the intercept; the variation around the linear fit is seen in the inset of Fig. \ref{fig:fig3}\subref{fig:fig3:a}.
	Measurements from this method are shown in Fig. \ref{fig:fig3}\subref{fig:fig3:a}, with measured loss as low as 1.35 dB/cm (TM) and 1.77 dB/cm (TE) at 405 nm and 2.89 dB/cm (TM) and 3.12 dB/cm (TE) at 371 nm.
	Measurements were taken for all three waveguide widths, and a strong dependence of loss on waveguide width was observed, shown in Fig. \ref{fig:fig3}\subref{fig:fig3:b}.
	
	The 600 nm width exhibited the lowest loss for $\lambda =$ 405, 419, and 458 nm, and the 500 nm width exhibited the lowest loss for $\lambda =$ 371 nm (the 600 nm wide waveguides were not tested at this wavelength). 
	The loss was uniformly lower for TM-like modes than TE-like modes, consistent with reduced field interaction with the sidewalls. 
	We note that previous work by our group in patterning alumina films using different tooling yielded losses between 5.8 and 6.8 dB/cm in the blue and UV, suggesting the process is robust \cite{gavin_ieee_sum}.
	
	\subsection{Ring Resonators}
	Ring resonator structures were fabricated with a 500 nm target waveguide width and 90 $\mu$m radius, where the TE-mode loss measured by the cutback method was 2.6 dB/cm at $\lambda =$ 405 nm. 
	From simulations TM-like modes are expected to experience high bending loss, and we measured no resonance with TM input. 

\begin{figure}[H]
			\centering
					{
				\phantomsubcaption\label{fig:fig4:a}
				\phantomsubcaption\label{fig:fig4:b}
				\phantomsubcaption\label{fig:fig4:c} 
					}
			\includegraphics{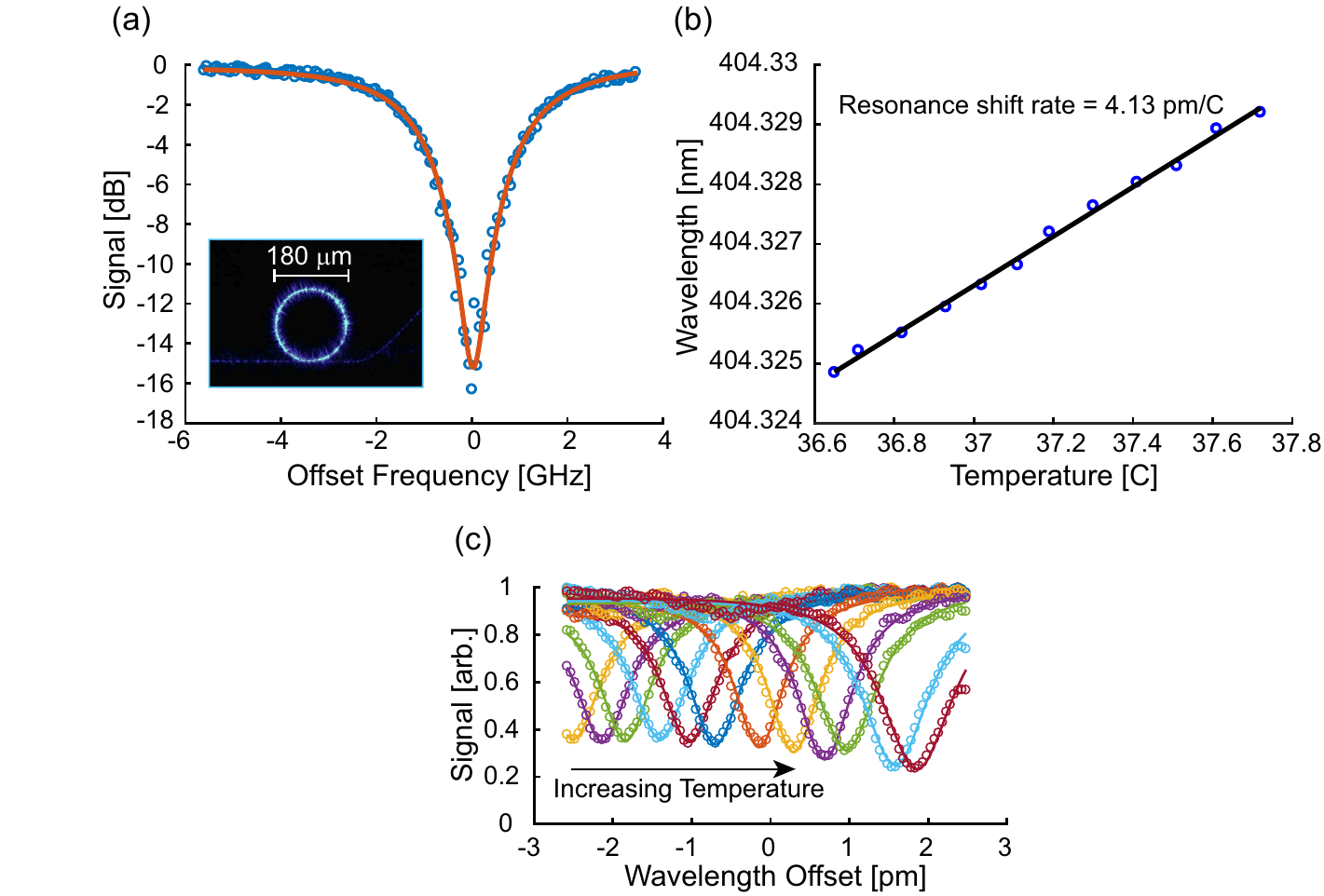}
			\caption{
				\subref*{fig:fig4:a} The experimental and fitted transmission of a slightly undercoupled ring, showing a 3 dB width of 2.7 GHz, corresponding to a loaded \emph{Q} of 275k. 
				Inset: Microscope image of ring under test. 
				\subref*{fig:fig4:b} The transmission minima wavelength as a function of temperature, giving the resonator thermo-optic shift. \subref*{fig:fig4:c} Undercoupled resonator transmission curves used to extract the thermal shift.
					}
			\label{fig:fig4}
		\end{figure}

	Measurements were taken with a tunable 405 nm Toptica DL100 external cavity diode laser, with mode-hop-free tuning achieved by a kHz-frequency triangle wave signal applied simultaneously to the laser diode current and output grating piezo. 
	A small portion of the laser output was directed into a Fabry-Perot etalon (Thorlabs SA200-3B, 1.5 GHz FSR, finesse $>$200) whose transmission was monitored simultaneously with the ring transmission. 
	Resonances of the ring were aligned to the center of the laser's tuning range by uniformly heating the resonator chip. 
	Transmitted light was measured with a silicon photodiode and the output monitored on an oscilloscope. 
	An example resonance is shown in Fig. \ref{fig:fig4}\subref{fig:fig4:a}, with the horizontal axis scaled using the Fabry-Perot fringes as a reference. 
	The quality factors were extracted by fitting data to the transmission curve of a Lorentzian oscillator normalized with respect to the incident power \cite{haus_text}.

	The resonator intrinsic quality factor and the coupling quality factor are not generally differentiable from a single measurement without knowledge of phase, so we measured rings with a variety of coupling gaps. 
	A 400 nm gap provided close to critical coupling; we extract an intrinsic \emph{Q} factor of 470,200 at $\lambda =$ 405 nm and a coupling \emph{Q} of 660,000. 
	Using ellipsometric index dispersion data and a 2D mode solver, we estimate a group index of $n_g =$ 1.648. 
	From this, the propagation loss in the resonator is approximately 2.35 dB/cm, calculated using
	
		\begin{align}
			\alpha & = 10 \log_{10}(e) \frac{2\pi n_g }{\lambda Q} \: [\text{dB/cm}]	\label{eq:lossQ}
		\end{align}
	
	By thermally tuning the chip without adjusting the laser frequency sweep span, we are able to measure the resonance shift due to the thermo optic (TO) effect. 
	TO tuning is a robust way of adjusting resonators to specific frequencies, or making phase modulators in materials lacking a direct electrical tuning mechanism.
	Using a linear fit of the resonance wavelength (Fig. \ref{fig:fig4}\subref{fig:fig4:b} and Fig. \ref{fig:fig4}\subref{fig:fig4:c}) we measure a shift of 4.13 pm/\degree C (-7.53 GHz/\degree C at $\lambda =$ 405 nm), which corresponds to an effective TO coefficient $1.68 \times 10^{-5}$ RIU/\degree C.
	This value is comparable to that of silicon nitride platforms operating at telecom wavelengths \cite{TO_coeff_measure}.

	An exact determination of the TO coefficient of bulk alumina requires information from two resonator modes. 
	However it is possible to estimate the value $dn_{core}/dT$ through Eq. \ref{eq:TOcoeff} using reported values for the SiO$_2$ coefficient and partial derivative sensitivity parameters calculated with a two dimensional mode solver. 
	
		\begin{align}
			\frac{d\lambda_{res}}{dT}  & = \frac{\lambda_{res}}{n_g}\left[\frac{\partial n_{eff}}{\partial n_{core}}\frac{dn_{core}}{dT}+ \frac{\partial n_{eff}}{\partial n_{clad}}\frac{dn_{clad}}{dT} \right] \label{eq:TOcoeff}
		\end{align}
	
	Taking the TO coefficient of oxide to be $1.0\times 10^{-5}$ [RIU/\degree C] \cite{to_coeff_sio2}, we estimate the bulk TO coefficient of ALD alumina to be $~2.75\times 10^{-5}$ [RIU/\degree C]. 
	This is consistent with previous ellipsometrically-derived TO values, reported to be in the range of   0.5-8$ \times 10^{-5}$[RIU/\degree C] at $\lambda=$ 633 nm for ALD alumina grown at 120\degree C \cite{alox_to_coeff}.
	We expect the TO coefficient to vary dependent on the deposition/anneal temperature and precursor hold-off times, particularly when the measurement temperature is near the deposition/anneal temperature.
		
\begin{figure}[H]
			\centering{
				\phantomsubcaption\label{fig:fig5:a}
				\phantomsubcaption\label{fig:fig5:b}
						}
			\includegraphics{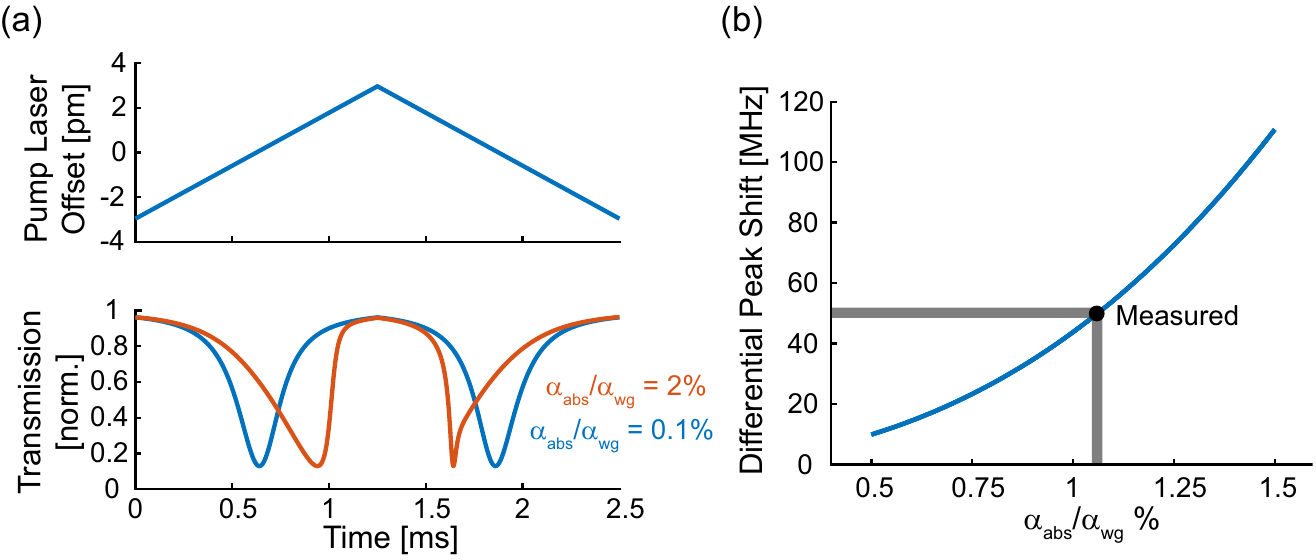}
			\caption{
				\subref*{fig:fig5:a} Top: The laser offset from the cold-cavity resonance as a function of time. 
				Bottom: Example modeled transmission curves with self-heating, for two absorption loss ratios, showing the asymmetry between upward and downward sweeps. 
				\subref*{fig:fig5:b} Modeled difference in transmission minimum between upward and downward sweep resonance.
				The semi-transparent overlay represents the spread in measured values.
			}
			\label{fig:fig5}
		\end{figure}

	High quality factors and strong TO coefficients can present a challenge in systems which can be limited by unwanted thermal tuning, such as refractive index sensors or those operating at high power, due to thermal self-instability \cite{vahala_thermal_instability}. 
	The self-heating mechanism is a function of absorptive losses, so an estimation of the absorption can be instrumental in system design. 
	It is possible to estimate the fraction of absorptive loss to total loss by examining the effect of the self-heating on the transmission line shape. 
	For positive TO coefficients, increases in the resonator temperature move the resonance toward longer wavelengths. 
	As a result, the measured transmission curve (and resonance wavelength) is different  when the laser wavelength sweep increases across the resonance versus decreases across the resonance.
	This behavior is illustrated in Fig. \ref{fig:fig5}\subref{fig:fig5:a}.
	We investigated this by reducing the frequency of our laser sweep to 400 Hz, fast enough to mitigate ambient fluctuations but much slower than the cavity thermal effects.
	The on-resonance intra-cavity power was calculated to be 12.5 mW, limited by coupling efficiency from fiber to waveguide and available laser power. 
	We compare our measurements to a model for the thermal self-heating, given in Ref. \onlinecite{vahala_thermal_instability}:

		\begin{align}
			C_p \Delta \dot{T}(t) &= I_p\frac{Q_{tot}}{Q_{abs}}  \frac{1}{\left( \frac{\lambda_p - \lambda_0(1+[\epsilon + \frac{1}{n_g} \frac{dn}{dT}]\Delta T(t))}{\Delta\lambda/2}\right)^2+1} - K\Delta T(t) \label{eq:ring_heating}
		\end{align}
		
	where $I_p$ is the loaded cavity power on resonance, $Q_{tot}$ is the loaded quality factor, $Q_{abs}$ is the quality factor in the case where the only source of loss is absorption, $\lambda_p$ is the pump laser wavelength, $\lambda_0$ is the cold cavity resonance wavelength, $\epsilon$ is the thermal expansion of the resonator material (from Ref. \onlinecite{alox_thermal_expansion}), $\Delta\lambda$ is the resonance full-width half-max (FWHM), $C_p$ is the thermal capacity of the system, and $K$ is the thermal conductance from the waveguide core to its surroundings, calculated using COMSOL. 
	The term $Q_{tot}/Q_{abs}$ acts as a fitting parameter, which using Eq. \ref{eq:lossQ} simplifies to $Q_{tot}/Q_{abs} = \alpha_{abs}/\alpha_{tot}$. $\alpha_{tot}$ includes coupling loss in the ring, so to extract the ratio of absorption loss to \emph{waveguide} loss we can multiply by the ratio of total loss to intrinsic loss $Q_{wg}/Q_{tot}$ which gives $\alpha_{abs}/\alpha_{wg}$.
	
	The resonance shift is measured by comparing the ring's resonance position, over one period of the applied triangle-wave sweep signal, to the reference Fabry-Perot fringes. 
	Resonance frequency is extracted from a Lorentzian fit of each individual resonance. 
	The induced asymmetry is measured  as $\Delta \nu = \nu_{res,\uparrow} - \nu_{res,\downarrow} \approx 50 $ MHz over five measurements.
	$\nu_{res,\uparrow}$ and $\nu_{res,\downarrow}$ are the optical frequencies corresponding to the transmission minimum when the laser frequency is increasing and decreasing across the resonance, respectively.
	This value is measurement limited by system stability and a sampling resolution of 17 MHz.
	By sweeping the parameter $\alpha_{abs}/\alpha_{wg}$ as shown in Fig. \ref{fig:fig5}\subref{fig:fig5:b} and comparing to the experimentally measured asymmetry, we estimate a conservative upper bound on the fraction of absorption loss $\alpha_{abs}/\alpha_{wg}\approx 1 \%$. 
	
	The observed immunity to absorption-induced heating and high quality factors make this platform well-suited for systems which require long-term stability and high powers, and suggests that significant improvements can be achieved by further reduction of the sidewall roughness.

	\section{Conclusion}
	Here we have demonstrated an amorphous aluminum oxide-based integrated photonics platform with low loss into the near UV, to our knowledge the lowest loss yet reported for fully-confined waveguides in the blue and ultraviolet.
	Furthermore, the fabrication tooling is commonly available in nanofabrication facilities, and the ALD process is both inexpensive and does not place restrictions on substrate material. 
	Propagation loss $<$3 dB/cm is measured at 371 nm, and $<$2 dB/cm at 405nm. Furthermore, ring resonators with intrinsic quality factors of 470,000 at 405 nm are measured, and the thermo-optic shift in our geometry is found to be $1.68 \times 10^{-5}$ [RIU/\degree C]. 
	Loss and thermal stability measurements suggest the loss is strongly dominated by scattering, instead of bulk material loss, so further reduction of loss at short wavelengths should be achievable with improvements in the fabrication process. 
	This opens routes to integrating technology reliant on short wavelengths, in particular biochemical sensors, nonlinear optics, and optical addressing of trapped ions, neutral atoms, and other quantum systems. 

	\section{Acknowledgments}
	This material is based upon work supported by the National Science Foundation Graduate Research Fellowship under Grant No. 1122374. This work was partially funded by NSF program ECCS-1408495. The work at Lincoln Laboratory was sponsored by the Assistant Secretary of Defense for Research and Engineering under Air Force contract number FA8721-05-C-0002. Opinions, interpretations, conclusions, and recommendations are those of the authors and are not necessarily endorsed by the United States Government.  
	The authors would like to thank Suraj Bramhavar at MIT Lincoln Laboratories and Amir Atabaki at MIT for many helpful discussions.

	\bibliographystyle{aipnum4-1}
	\bibliography{AlOxWaveguides}

\end{document}